\documentclass[%
 reprint,
 amsmath,amssymb,
 aps,
 prd,
 longbibliography,
 floatfix,
 superscriptaddress
]{revtex4-1}
\usepackage{amsfonts, relsize, color}
\usepackage{graphics}
\usepackage{graphicx}
\usepackage{subfig}
\usepackage{hyperref}
\usepackage{amssymb}
\usepackage{mathrsfs}
\usepackage{siunitx}
\DeclareSIUnit{\wtpercent}{wt\%}
\usepackage{amsmath}
\usepackage{diagbox}
\usepackage[version=4]{mhchem}

\usepackage{bm}

\makeatletter
\renewcommand{\fnum@figure}{FIGURE. \thefigure}
\makeatother

\newcommand{\be}{\begin{equation}}
\newcommand{\ee}{\end{equation}}
\newcommand{\datasetname}{OMDB-GAP1}
\newcommand{\bestEnsembleMAE}{0.388}

\usepackage[normalem]{ulem}

\begin{document}
\title{Band gap prediction for large organic crystal structures with machine learning}
\author{Bart Olsthoorn}
\email{bartol@kth.se}
\affiliation{Nordita, KTH Royal Institute of Technology and Stockholm University, Roslagstullsbacken 23, SE-106 91 Stockholm, Sweden}
\author{R. Matthias Geilhufe}
\email{geilhufe@kth.se}
\affiliation{Nordita, KTH Royal Institute of Technology and Stockholm University, Roslagstullsbacken 23, SE-106 91 Stockholm, Sweden}
\author{Stanislav S. Borysov}
\email{stabo@dtu.dk}
\affiliation{Technical University of Denmark, DTU, 2800 Kgs. Lyngby, Denmark}
\author{Alexander V. Balatsky}
\affiliation{Nordita, KTH Royal Institute of Technology and Stockholm University, Roslagstullsbacken 23, SE-106 91 Stockholm, Sweden}

\date{June 7, 2019}

\begin{abstract}
Machine learning models are capable of capturing the structure-property relationship from a dataset of computationally demanding \textit{ab initio} calculations. In fact, machine learning models have reached chemical accuracy on small organic molecules contained in the popular QM9 dataset. At the same time, the domain of large crystal structures remains rather unexplored. Over the past two years, the Organic Materials Database (OMDB) has hosted a growing number of electronic properties of previously synthesized organic crystal structures. The complexity of the organic crystals contained within the OMDB, which have on average 82 atoms per unit cell, makes this database a challenging platform for machine learning applications. In this paper, we focus on predicting the band gap which represents one of the basic properties of a crystalline material. With this aim, we release a consistent dataset of \num{12 500} crystal structures and their corresponding DFT band gap freely available for download at \url{https://omdb.mathub.io/dataset}. We run two recent machine learning models, kernel ridge regression with the Smooth Overlap of Atomic Positions (SOAP) kernel and the deep learning model SchNet, on this new dataset and find that an ensemble of these two models reaches mean absolute error (MAE) of \SI{\bestEnsembleMAE}{\electronvolt}, which corresponds to a percentage error of \SI{13}{\percent} on the average band gap of \SI{3.05}{\electronvolt}. Scaling properties of the prediction error with the number of materials used for model training indicate that a prohibitively large number of materials would be required to reach chemical accuracy. It suggests that, for the large crystal structures, machine learning approaches require development of new model architectures which utilize domain knowledge and simplifying assumptions. The models also provide chemical insights into the data. For example, by visualizing the SOAP kernel-based similarity between the crystals, different clusters of materials can be identified, such as organic metals or semiconductors. Finally, the trained models are employed to predict the band gap for \num{260092} materials contained within the Crystallography Open Database (COD) and made available online so the predictions can be obtained for any arbitrary crystal structure uploaded by a user.
\end{abstract}

\maketitle

\section{Introduction}
Many properties of a crystalline material, such as electric conductivity or optical absorption spectrum, are primarily governed by its electronic structure stemming from the underlying quantum mechanical nature of the electrons. The ability to find or even design materials with target functional properties is of great importance to sustain current technological progress. Although ever-growing computational resources and better algorithms have significantly accelerated this search, the combinatorial complexity of the problem requires new approaches to be employed.

In the recent decades, the amount of scientific data collected has facilitated the emergence of new data-driven approaches in the search for novel functional materials. Scientific data has been made accessible in terms of a multitude of online databases, e.g., for crystal structures \cite{COD1,COD2,CSD,ICSD}, electronic structures and materials properties \cite{materials_project,Borysov2017,draxl2018nomad,gurka,pubchemqc}, enzymes and pharmaceutics \cite{drugbank,schomburg2002brenda}, or superconductors \cite{supercon,wimbush2017public}. In contrast to pure data-mining approaches, which focus on extracting knowledge from existing data \cite{borysov2018online,geilhufe2018towards,Suram2015}, machine learning approaches try to predict target properties directly, where a highly non-linear map between a crystal structure and its functional property of interest is approximated. In this context, machine learning offers an attractive framework for screening large collections of materials. Having an accurate machine learning model at hand can tremendously accelerate the identification of novel functional materials, as the prediction of the property of interest for a given crystal structure bypasses computationally expensive modelling based on {\it ab initio} methods. 

There has been a growing interest in developing interpretable and efficient machine learning models for materials science and quantum-chemical systems \cite{pmlr-v70-gilmer17a,Schtt2017,Brockherde2017,Stanev2018}. It has been reported that machine learning models have reached chemical accuracy on prediction tasks for various datasets such as the popular QM9 dataset with small organic molecules~\cite{De2016,Schtt2018}. Meanwhile, the Materials Project is often used to test the predictive power of models trained on mostly inorganic crystal structures~\cite{PhysRevLett.120.145301,Schtt2018}. However, the Organic Materials Database (OMDB) provides a more challenging task: the prediction of properties for complex and lattice periodic organic crystals with a much larger number of atoms in the unit cell.

Here, an important and difficult part is to numerically represent molecules or crystal structures in a way suitable for machine learning, which has been the topic of many works \cite{PhysRevLett.108.058301,Hansen2015,2015arXiv150307406F,1704.06439}. These representations should incorporate the known invariances of the molecule or crystal such as translational or rotational invariance and the choice of the unit cell. Recently, there have been successful approaches that learn these representations directly by incorporating them into the model architecture \cite{pmlr-v70-gilmer17a,Schtt2017,Schtt2018}.

Organic materials have attracted a lot of attention with respect to spintronic devices \cite{Spintronics1, Spintronics2, Spintronics3} and magnon spintronics \cite{liu2018organic}, molecular qubits \cite{qubit1,qubit2}, spin-liquid physics \cite{spinliquid1,spinliquid2,spinliquid3}, and, last but not least, organic LEDs and solar cells \cite{oled1,photov}. For the latter two, the size of the band gap, i.e., the energy distance between the lowest unoccupied and the highest occupied electronic states, plays a significant role. Combining the design of a material with the optimal band gap regime with the soft elastic properties inherent to organic materials opens a path towards the engineering of novel flexible electronic devices.

Given the technological importance of band gap predictions and the rapid progress in the field of machine-learning-based materials design, we present a newly released and freely available band gap dataset (\datasetname{}) at \url{https://omdb.mathub.io/dataset}, containing the band gaps of \num{12500} three-dimensional organic molecular crystals calculated using density functional theory (DFT). We discuss the performance of recent machine learning models and crystal structure representations, namely, kernel ridge regression based on the Smooth Overlap of Atomic Positions (SOAP) kernel~\cite{PhysRevB.87.184115} and the deep learning model SchNet~\cite{Schtt2018}, to provide a benchmark for the state-of-the-art given this rather small, but complex dataset. The trained models presented throughout this paper are publicly available via a web interface at \url{https://omdb.mathub.io/ml} (see Appendix~\ref{appendix:web_interface}).

We have previously reported on the Organic Materials Database (OMDB) and its web interface~\cite{Borysov2017}. In this paper we model, for the first time, the structure-property relationship with machine learning for organic crystals more complex than any dataset currently available. We benchmark two currently-available state-of-the-art methods that have been used for small molecules and simple inorganic crystals, and evaluate their performance on this new challenging dataset.

The rest of the paper is structured as follows. To provide a more detailed description of our new publicly available band gap dataset (\datasetname{}), we review the implementation of the OMDB in Section 2 and provide additional information about the statistics of the \datasetname{} dataset. In Results and Discussion, we apply the two machine learning methods mentioned above and discuss their performance on this dataset. We also show an application example of large-scale material screening with the trained models. Additional information about the machine learning models, hyperparameter optimization, and the newly developed web interface can be found in the appendix. We summarize the paper in Conclusion.

\section{OMDB-GAP1 -- A new dataset}
Table~\ref{tab:datasets} lists the most common datasets currently used for statistical modeling of functional properties. Small organic molecules are well accounted for with the QM9 dataset \cite{Ruddigkeit2012,ramakrishnan2014quantum}, which has enabled many pioneering studies. The new dataset introduced here, \datasetname{}, comes with two advantages over existing datasets. First, the dataset is consistent, i.e., all the calculations were performed within the same DFT setup. Secondly, the crystal structures are, on average, much larger than other available datasets. The unit cell size ranges from 7 to 208 atoms, with an average of 82 atoms. The \datasetname{} dataset is available for download at \url{https://omdb.mathub.io/dataset}.
\begin{table}[b]
    \centering
    \caption{Comparison of the existing datasets currently used for machine learning approaches with \datasetname{}. $\overline{N}$ indicates the average number of atoms in the molecule or unit cell. Consistency means that all data comes from identical computational setup.}
    \begin{ruledtabular}
    \begin{tabular}{ccccc}
        \textbf{Name} & \textbf{Size} & \textbf{Type} & $\overline{N_a}$ & \textbf{Consistent} \\
        QM9 & \num{133885} & Organic molecules & 18 & \checkmark\\
        Mat. Pro. & \num{53340} & Crystals & 27\footnote{Materials project is quota limited so $\overline{N_a}$ is estimated from the histogram in the supplementary material of~\cite{PhysRevLett.120.145301}.} & $\text{\sffamily X}$\\
        \datasetname{} & \num{12500} & Organic crystals & 82 & \checkmark \\
    \end{tabular}
    \end{ruledtabular}
    \label{tab:datasets}
\end{table}

\begin{figure}[t]
\subfloat[\label{fig:bandgap_histogram}]{\includegraphics[]{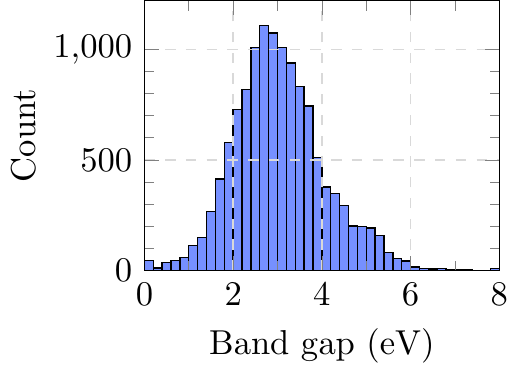}}\\\vspace{-.5cm}
\subfloat[\label{fig:spacegroups_count}]{\includegraphics[]{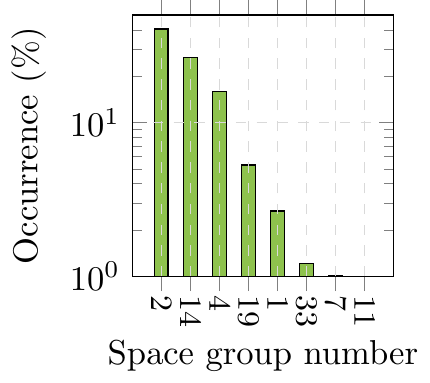}}
\subfloat[\label{fig:species_count}]{\includegraphics[]{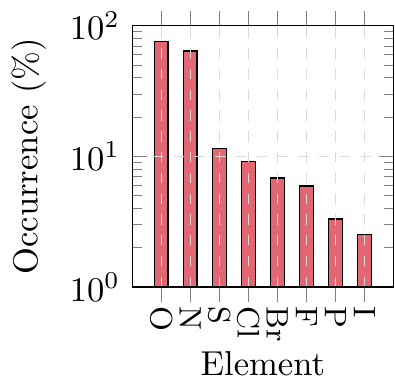}}
\caption{Descriptive statistics of the \datasetname{} dataset. (a) Distribution of all \num{12500} band gaps with population statistics: $\text{mean}=3.05$, $\text{median}=2.96$, $\text{standard deviation}=1.03$. The distribution exhibits the form of the Wigner-Dyson distribution with the estimated parameters: $\sim x^{4.61}e^{-0.28 x^2}$ (b) The most common space groups. (c) The most common elements, excluding C and H, which occur in all the structures.}
\label{fig:dataset_properties}
\end{figure}

The dataset presented here is a subset of all the calculations contained in the OMDB, which were discussed in detail in Ref.~\cite{Borysov2017}. From all the materials stored within the OMDB, only materials with a calculated magnetic moment of less than $10^{-4}$~$\mu_B$ are selected. In total, the \datasetname{} dataset contains band gap information for \num{12500} materials. The dataset comprises 65 elements, with the heaviest element being Uranium, and spans 69 space groups. Figure~\ref{fig:spacegroups_count} and Figure~\ref{fig:species_count} show the most common space groups and atomic elements, respectively. All contained band gaps were calculated in the DFT framework by applying the Vienna ab initio simulation package VASP \cite{vasp1,vasp2}, which is based on the pseudopotential projector augmented-wave method. This approach is particularly suitable to treat the sparse unit cell structures immanent to organic molecular crystals. The initial structural information of the investigated materials was taken from the Crystallography Open Database (COD)~\cite{COD1,COD2}. The energy cut-off was chosen to be the maximum of the specified maxima for the cut-off energies within the POTCAR files (precision flag ``normal''). For the integration in $\vec{k}$-space a $6\times6\times6$ automatically generated, $\Gamma$-centered mesh was used.

Organics were reported to contain strongly correlated electrons as well as intermolecular van-der-Waals interactions that usually require the application of advanced exchange-correlations functionals, such as Meta-GGAs or hybrid functionals as well as van-der-Waals interactions \cite{organicsDFT1,organicsDFT2,organicsDFT3}. However, to obtain a statistically relevant dataset while keeping a reasonable computational demand, we chose the generalized gradient approximation according to PBE~\cite{perdew1996}. The distribution of the calculated band gaps exhibits the form of the Wigner-Dyson distribution ($\sim x^\alpha \exp(-\beta\ x^2)$) and is shown in Figure~\ref{fig:bandgap_histogram}. Choosing PBE can introduce a systematic error into the calculations (e.g. changing the position of the mean and the size of the variance), however, it does not affect the overall statistical properties of the dataset (shape of the band gap distribution). In particular, the latter is important for the development of machine learning models, which, once they reach chemical accuracy, can be applied to an improved dataset whenever present.

Typically, the computational demand for DFT calculations tremendously increases with the number of atoms $N_a$ in the unit cell. Even though implementations with $\mathcal{O}\left(N_a\right)$ scaling were reported~\cite{guerra1998,zeller2008}, commonly used DFT codes scale with $\mathcal{O}\left(N_a^2\log N_a\right)$ up to $\mathcal{O}\left(N_a^3\right)$~\cite{kresse1996efficiency}. This represents a serious problem for \emph{ab initio} modeling of organic crystal structures. Next to the number of atoms, also other parameters like the energy cut-off for the plane-wave expansion influence the computing time. For the present dataset, the calculations were running for $\approx 60$ core hours for an average material, leading to a total estimate of 750k core hours for the entire dataset.

\section{Results and Discussion}
\subsection{Machine learning for organic crystals}
\begin{figure}[t]
\subfloat[\label{fig:soap_predictions}]{\includegraphics[]{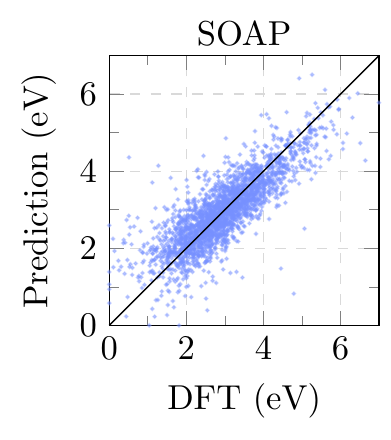}}
\hspace{0.2cm}
\subfloat[\label{fig:schnet_predictions}]{\includegraphics[]{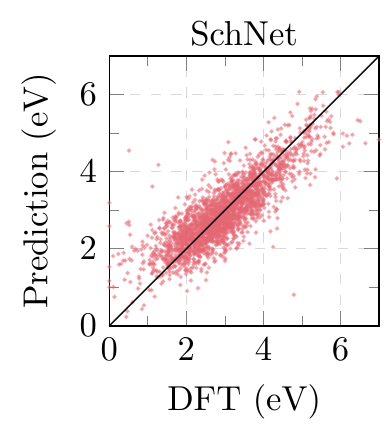}}\\\vspace{-.5cm}
\subfloat[\label{fig:ensemble_predictions}]{\includegraphics[]{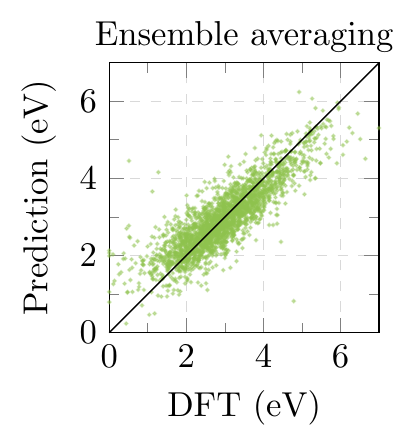}}
\caption{Band gap predictions for the \num{2500} test materials for the models (a) SOAP (MAE \SI{0.430}{\electronvolt}), (b) SchNet (MAE \SI{0.415}{\electronvolt}) and the (c) SOAP/SchNet ensemble (MAE \SI{\bestEnsembleMAE}{\electronvolt}).}
\label{fig:predictions}
\end{figure}

For a prediction of the electronic band gap based on machine learning we discuss two recent models: kernel ridge regression with the SOAP kernel~\cite{PhysRevB.87.184115} and the deep learning model SchNet~\cite{Schtt2018}. A brief introduction to both models is given in Appendix~\ref{appendix:methods}. As an input, we use the atom positions and corresponding atomic numbers. We train the models using \num{10000} materials (training set) and evaluate their performance using Mean Absolute Error (MAE) and Root Mean Squared Error (RMSE) for the remaining \num{2500} materials (test set).

Table~\ref{tab:summary_performance} and Figure~\ref{fig:predictions} summarize the best performance of each model after hyperparameter optimization (more details given in the appendix). The ``Constant'' model refers to using the  mean (for RMSE) and median (for MAE) values from the training set as the most basic baselines to calculate, respectively. Although SchNet is more accurate (MAE \SI{0.415}{\electronvolt}) than SOAP (MAE \SI{0.430}{\electronvolt}), it overestimates the band gap of the metals and small-gap materials (Fig.~\ref{fig:schnet_predictions}). For SOAP, the MAE on the training data is \SI{0.342}{\electronvolt}, indicating that the model is overfitting slightly. In contrast, the MAEs for SchNet on the training and validation set are \SI{0.274}{\electronvolt} and \SI{0.397}{\electronvolt}, respectively. The deep learning model SchNet has more free parameters than SOAP, which makes it more prone to overfitting.
The result can be slightly improved by considering an ensemble average prediction \cite{murphy2012machine}, leading to a MAE of \SI{\bestEnsembleMAE{}}{\electronvolt} and an RMSE of \SI{0.519}{\electronvolt}, indicating that the errors of both models are slightly uncorrelated. Considering a mean band gap of \SI{3.05}{\electronvolt} as shown in Figure~\ref{fig:bandgap_histogram}, this corresponds to a percentage error of $\approx$\SI{13}{\percent}. Whether this performance is accurate enough to screen for materials and how many complementary computationally expensive DFT calculations are required depends on the application domain. For example, efficiency of solar cells can be estimated using the Shockley-Queisser (SQ) limit for a single p-n junction. For the band gap of \SI{1.34}{\electronvolt}, it gives the maximum performance corresponding to \SI{33.7}{\percent}. In this case, the MAE of \SI{0.388}{\electronvolt} would correspond to \SI{5.1}{\percent} performance decrease \cite{Rhle2016}.
 
\begin{figure}[b]
    \centering
    \includegraphics[]{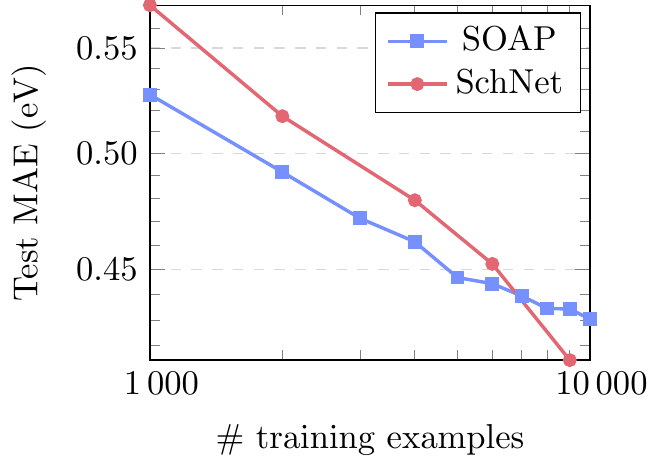}
    \caption{Benchmark of SOAP and SchNet on the OMDB band gap dataset (\datasetname{}). SOAP is used with an average kernel and with $r_c=\SI{4}{\angstrom},n=8,l=6$. SchNet is used with $T=3,F=64$ and a validation set of \num{1000} materials is used for early stopping.}
    \label{fig:best_benchmark}
\end{figure}

\begin{table}[b]
    \centering
    \caption{Performance of different models on the test set (complete training set is used for training). The ``constant'' model corresponds to predicting the median and mean of the training set for MAE and RMSE, respectively.}
    \begin{ruledtabular}
    \begin{tabular}{cccc}
         & MAE (eV) & RMSE (eV)\\
        \hline\\
        Constant & 0.797 & 1.035\\
        SOAP & 0.430 & 0.576\\
        SchNet & 0.415 & 0.554\\
        Ensemble averaging & \textbf{\bestEnsembleMAE{}} & \textbf{0.519}
    \end{tabular}
    \end{ruledtabular}
    \label{tab:summary_performance}
\end{table}

\begin{figure*}[t]
    \centering
    \includegraphics{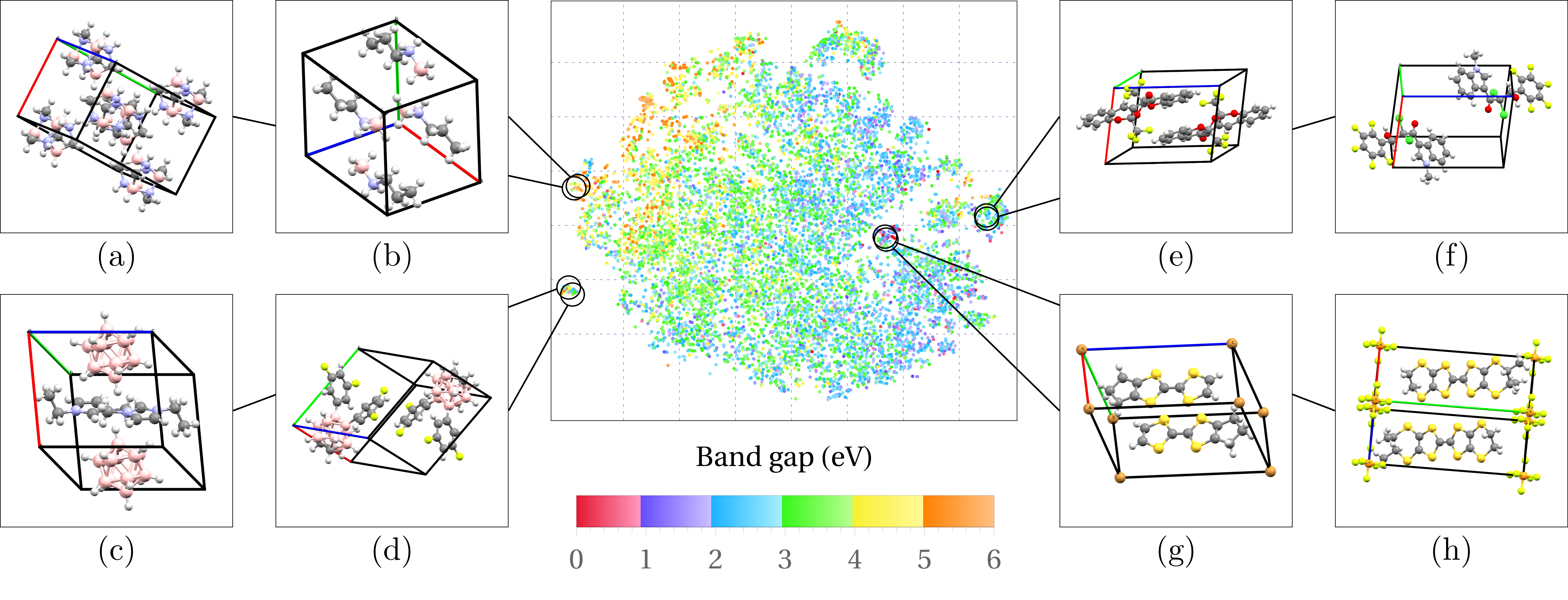}
    \caption{Dimensionality-reduced visualization of the crystal structures with the t-SNE algorithm~\cite{maaten2008visualizing} with perplexity 50. The distance between two structures is determined by the average SOAP kernel ($r_c=\SI{4}{\angstrom},n=8,l=6$). The PBE band gap is color-coded for each structure. Regions with both similar structure and band gap can be recognized.  
    Structures related to chemical hydrogen storage devices appear in the vicinity of (a) OMDB-ID 36754 (\ce{C4 H14 B4 N4}) and (b) OMDB-ID 22958 (\ce{C3 H12 B N}).
    Structures with similar Boron-containing clusters form the island including (c) OMDB-ID 4085 (\ce{C12 H34 B12 N4}) and (d) OMDB-ID 26557 (\ce{C14 H16 B10 F4}).
    Molecular crystals are present in in the region of (e) OMDB-ID 1637 (\ce{C21 H10 F6 O4}) and (f) OMDB-ID 4532 (\ce{C18 H10 Cl2 F5 N O2}).
    A molecular metal (g) OMDB-ID 16923 (\ce{C18 H16 Br S8}) and a structurally similar organic superconductor (h) OMDB-ID 33839 (\ce{C24 H24 F6 P S16}) appear in the same zero band gap region.
    Online interactive version (with zoom) available at \url{https://omdb.mathub.io/dataset/OMDB-GAP1/interactive}.}
    \label{fig:t-SNE-SOAP}
\end{figure*}

Figure~\ref{fig:best_benchmark} shows the mean absolute error (MAE) decrease with the number of training examples $N$ on a log-log scale, revealing a roughly linear trend. For SOAP and SchNet a power law is fitted to $0.965N^{-0.087}$ and $1.507N^{-0.140}$, respectively. We use this approximation to estimate the number of materials necessary to reach a certain level of predictive power. For example, if we aim for a MAE of \SI{0.1}{\electronvolt}, which corresponds to a percentage error of \SI{3}{\percent} on the average band gap of \SI{3.05}{\electronvolt}, the upper bound for SchNet is approximately 267M materials, which is far beyond the scale of available band gap data for organic molecular crystals (e.g., the COD contains crystallographic information for $\approx$ \num{325000} organic crystals). As this number is prohibitively large with current computational resources, it indicates the necessity for more powerful machine learning models and/or simplifying assumptions which incorporate domain knowledge to reach this level of accuracy. Meanwhile, the current accuracy can be used to reduce the search space for high-throughput screening with DFT. 

The model based on the average SOAP kernel has fewer free parameters than SchNet. Therefore, it is expected that SOAP outperforms SchNet for a small number of training examples, where SchNet is more likely to overfit. However, training of SchNet has linear complexity with the number of training examples, whereas the complexity is cubic for SOAP. In practice, this means SchNet can be trained on a larger set of materials.

To visualize the relationship between the \datasetname{} crystal structures and their band gap, we used the learned SOAP kernel to measure the structural similarity of two materials $A$ and $B$ as $D(A,B)^2=K(A,A)+K(B,B)-2K(A,B)$~\cite{De2016}. This distance measure is used by a dimensionality reduction algorithm t-SNE~\cite{maaten2008visualizing} to construct a two-dimensional projection of the high-dimensional crystal structure data. The results shown in Figure~\ref{fig:t-SNE-SOAP} exhibit a clear gradient in the band gap size from materials clustered in the upper left corner (large band gap) to materials in the lower right corner (small band gap). We furthermore picked a few examples where we illustrate a selection of crystal structures visualized with the program {\it Mercury}~\cite{Macrae:ks5091}. For example, structures related to chemical hydrogen storage devices appear in the vicinity of each other, such as Bis-BN Cyclohexane (Figs.~\ref{fig:t-SNE-SOAP}(a) with \SI{4.7}{\wtpercent} hydrogen storage capability\cite{Chen2014} and Propylamine--borane (Figs.~\ref{fig:t-SNE-SOAP}(b). Similarly, crystals with Boron-containing clusters, which have applications in the pharmaceutical industry~\cite{Lenikowski2016}, are shown in Figs.~\ref{fig:t-SNE-SOAP}(c) and (d), and pure molecular crystals can be identified in Figs.~\ref{fig:t-SNE-SOAP}(e) and (f). The 36 metals present in the \datasetname{} shown in Fig.~\ref{fig:t-SNE-SOAP}(g) appear close to the structurally similar organic superconductor $\beta\text{-(meso-DMBEDT-TTF)}_2\text{PF}_6$ (transition temperature $T_c=\SI{4.2}{\kelvin}$ \cite{B409631B}) shown in Fig.~\ref{fig:t-SNE-SOAP}(h).
\begin{figure}[b]
    \centering
    \includegraphics{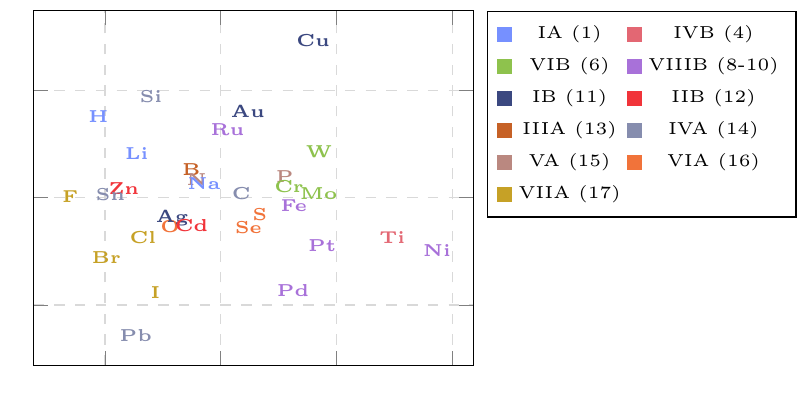}
    \caption{Principal component analysis (PCA) of the learned embeddings $\bm{a}_Z$ ($F=64$-dimensional) in SchNet ($T=1$, 9000 training examples). Elements that occur in fewer than 35 materials are ignored. Groups of the periodic table denoted by CAS numbering can be recognized.}
    \label{fig:schnet_embeddings}
\end{figure}

The dimensionality-reduced visualization where the functional property (e.g. band gap) is color-coded provides an intuitive way to navigate the structure-property landscape of organic crystals. With that it provides a basic approach to narrow down the search space to identify materials with a specified target functionality, e.g., materials for hydrogen storage or superconductors. In addition to the previous examples, by zooming in on certain areas it is possible to find structures forming a line that represents a study of the effect of pressure on a crystal structure. Other regions of interest can be explored using the online interactive version at \url{https://omdb.mathub.io/dataset/OMDB-GAP1/interactive}, which also provides a zoom functionality.

The way SchNet model calculates its final prediction as an average of individual atomic contribution makes it possible to identify structure-property relationships for separate atoms. The atom embeddings $\bm{a}_Z$ (with dimensionality $F$) are initialized randomly and updated while training the network. The learned embeddings can be studied in a dimensionality-reduced picture as well, which shows that SchNet has learned similarity between certain atom types as in the periodic table (Figure~\ref{fig:schnet_embeddings}). Generally speaking, elements that have a similar local environment and contribute similarly towards the final prediction are expected to have similar embeddings, but the exact way chemical information stored in the embeddings while training is non-trivial. We observe that the groups such as VIIA and VIB form clusters. Note that extracting the periodic table from machine learning models is also possible in the case of SOAP with a chemical environment kernel~\cite{2018arXiv180700236W}. Alternatively, it is also possible to use only compositional information to create atom embeddings that can reconstruct the periodic table~\cite{Zhou201801181}.

\subsection{High-throughput screening of the COD Database}
The previously trained models are able to make band gap predictions for millions of organic crystal structures and narrow the search space down where computationally expensive conventional methods such as DFT can subsequently be used.
As a demonstration, we employ the trained models to calculate the band gap for the materials contained in the Crystallography Open Database (COD). The database contains \num{399020} Crystallographic Information Files (CIF), of which \num{307013} are organic (i.e. include C and H) and consist of the 65 elements present in \datasetname{}. The COD has an average of \num{317} atoms in the unit cell, significantly larger than the average of 82 atoms within our dataset. To eliminate extremely large structures we only consider structures with the maximum of 500 atoms in the unit cell, leading to \num{260092} structures. 

\begin{figure}[t]
\centering
\includegraphics[]{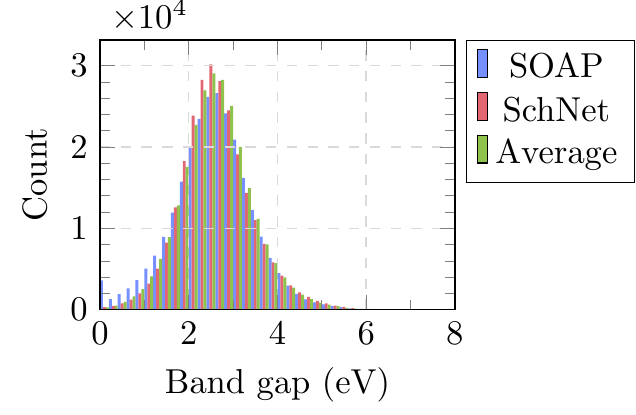}
\caption{Distribution of \num{260092} COD band gap predictions. The distribution exhibits the form of the Wigner-Dyson distribution with estimated parameters $\sim x^{5.31}e^{-0.41 x^2}$, $\sim x^{6.39}e^{-0.51 x^2}$ and $\sim x^{6.03}e^{-0.48 x^2}$ for SOAP, SchNet and the ensemble average, respectively.}
\label{fig:cod_histogram}
\end{figure}

Figure~\ref{fig:cod_histogram} shows the distribution of band gap predictions which is similar to Figure~\ref{fig:bandgap_histogram}. The COD predictions are available to download from \url{https://omdb.mathub.io/dataset}. As a proof of concept, we search for predicted solar cells around the Shockley-Queisser (SQ) limit \SI[separate-uncertainty = true]{1.34\pm0.05}{\electronvolt} of the SOAP/SchNet average ensemble predictions and find 3343 candidate materials (excluding materials present in \datasetname{}). This represents \SI{1.3}{\percent} of the initial search space on which more accurate high-throughput DFT calculations can be used for verification. Table~\ref{tab:solar_cells} lists a random selection of candidate materials (with $N<200$) for solar cells identified by the SOAP/SchNet average ensemble (i.e. materials with a predicted band gap within the range of \SI[separate-uncertainty = true]{1.34\pm0.05}{\electronvolt}. Note that this search does not take the band dispersion into account and whether a band gap is direct or indirect. A more extensive study focused on organic solar cells will be the topic of future work.

\begin{table}[b]
    \centering
    \caption{Random selection of 15 candidate materials with $N\leq 200$ for solar cells identified by SOAP/SchNet average ensemble. The predicted band gaps are listed as well as the number of atoms in the unit cell ($N_a$) and the space group number (SG).}
    \begin{ruledtabular}
    \begin{tabular}{lcccc}
        Compound & N & COD-ID & SG & Band gap (eV)\\
        \hline
            \ce{C34H40Co3N16O8} & 101 & 7033509 & 2 & 1.387\\
            \ce{C32H40Cl8Mn4N16O8} & 108 & 7013640 & 15 & 1.293\\
            \ce{C52H36N20O28} & 136 & 7219834 & 14 & 1.336\\
            \ce{C40H40Cu4F24N8O8W4} & 128 & 2006452 & 14 & 1.301\\
            \ce{C84H60Cu4N12O20} & 180 & 2215189 & 14 & 1.363\\
            \ce{C36H40Cu4I4N20O4} & 108 & 7009105 & 14 & 1.341\\
            \ce{C70H40Cl14N6O6Ru2} & 138 & 7027581 & 2 & 1.340\\
            \ce{C48H32Cu4I4N8} & 96 & 2227752 & 19 & 1.293\\
            \ce{C52H44AlN4O2} & 103 & 4331896 & 2 & 1.389\\
            \ce{C28H32N4Ni2O12} & 78 & 2227456 & 14 & 1.352\\
            \ce{C48H60Ba2N8O32} & 150 & 2012449 & 14 & 1.356\\
            \ce{C94H64Co4N24O6S8} & 200 & 2239446 & 4 & 1.340\\
            \ce{C34H42FeN4NiS6} & 88 & 4322934 & 2 & 1.346\\
            \ce{C56H48N8O14Zn2} & 128 & 7229269 & 2 & 1.381\\
            \ce{C32H32Cl4N8Ni2O8} & 86 & 2235069 & 14 & 1.318
    \end{tabular}
    \end{ruledtabular}
    \label{tab:solar_cells}
\end{table}

\section{Conclusion}
We present the new \datasetname{} dataset which contains DFT bandgaps for \num{12500} organic crystal structures, which span 69 space groups and 65 elements, to facilitate data-driven approaches in materials science. The structures have on average 82 atoms per unit cell which represents a challenging benchmark for machine learning models comparing to the existing materials datasets.
We train two recent machine learning models, kernel ridge regression with the Smooth Overlap of Atomic Positions (SOAP) kernel and the deep learning model SchNet, on this new dataset and find that an ensemble of these two models reaches mean absolute error (MAE) of \SI{\bestEnsembleMAE}{\electronvolt}, which corresponds to a percentage error of \SI{13}{\percent} on the average band gap of \SI{3.05}{\electronvolt}. Scaling properties of the prediction error with the number of materials used for model training indicate that a prohibitively large number of materials would be required to reach chemical accuracy. It suggests that, for the large crystal structures, machine learning approaches require development of new model architectures which utilize domain knowledge and simplifying assumptions. 

The models can also provide chemical insights into the data. For example, by visualizing the SOAP kernel-based similarity between the crystals, different clusters of materials can be identified, such as organic metals or semiconductors. Finally, the trained models are employed to predict the band gap for \num{260092} materials contained within the Crystallography Open Database (COD) and made available online (see Appendix~\ref{appendix:web_interface}) so the predictions can be obtained for any arbitrary crystal structure uploaded by a user.

\section*{Data Availability}
The datasets are available from \url{https://omdb.mathub.io/dataset}, including both the OMDB-GAP1 dataset and the COD predictions. To facilitate the training of the SchNet model on this dataset, our preprocessing script is available in the official SchNetpack repository at \url{https://github.com/atomistic-machine-learning/schnetpack/tree/master/src/scripts/schnetpack_omdb.py}.

\section*{Acknowledgement}\label{acknowledgements}
We are grateful for support from the Swedish Research Council Grant No.~638-2013-9243, the Knut and Alice Wallenberg Foundation, the VILLUM FONDEN via the Centre of Excellence for Dirac Materials (Grant No. 11744), the European Research Council under the European Union’s Seventh Framework Program (FP/2207-2013)/ERC Grant Agreement No.~DM-321031, and the Marie Sklodowska-Curie grant agreement no.~713683 (COFUNDfellowsDTU). The authors acknowledge computational resources from the Swedish National Infrastructure for Computing (SNIC) at the Center for High Performance Computing (PDC) and the High Performance Computing Center North (HPC2N) and the Uppsala Multidisciplinary Center for Advanced Computational Science (UPPMAX).

\section*{Author Contributions}
All authors designed the study. B.O., S.S.B. and R.M.G. prepared the dataset and performed the machine learning study. R.M.G. performed the DFT calculations. All authors analyzed the results, wrote, and revised the manuscript.

\section*{Competing Interests}
The authors declare no competing interests.

\appendix

\section{Machine Learning Models}\label{appendix:methods}
From a machine learning point of view, band gap prediction represents a regression task, which aims to find a non-linear relation $y=f(\bm{x})$ between an input vector $\bm{x}$ (commonly referred to as ``features'') and an output $y$ (commonly referred to as ``target''). The information about a molecule or crystal structure is usually represented as atomic positions and atomic numbers. A common strategy is to construct a descriptor $\bm{x}$ that encodes this information in a fixed-size vector. Depending on the context in which the descriptors are used, certain properties such as translation, rotational, atom index invariance and differentiability are important \cite{PhysRevLett.108.058301,2015arXiv150307406F}. This task is difficult since equivalent crystal structures can easily lead to distinct representations, causing inconsistent data for numerical methods. Many different descriptors have been introduced in the literature, such as the Coulomb matrix~\cite{PhysRevLett.108.058301}, Bag-of-Bonds~\cite{Hansen2015}, Sine Matrix~\cite{2015arXiv150307406F} and MBTR~\cite{1704.06439}. An alternative strategy is to skip the intermediate step of fixed-size descriptors, which often involve ad-hoc decisions, and directly define the kernel between two structures as in the Smooth Overlap of Atomic Positions (SOAP) kernel, which has shown superior performance on a variety of problems~\cite{De2016,Jger2018}.

In this paper, two recent machine learning models are evaluated on the new dataset: SOAP~\cite{PhysRevB.87.184115} and SchNet~\cite{Schtt2018}. This provides a baseline estimate of the predictive power of current models for the presented dataset. In this study, the atomic numbers and positions are available, but there exist alternative approaches if only the composition of a crystal is known, such as the automatically generated features from Magpie~\cite{Ward2016}.

The models studied here can be used for a fast screening of desired material properties, after which traditional calculations can be used to verify the results. Additionally, the considered models are differentiable with respect to atom positions, which means that the output property can be tuned through gradient descent. When a model is trained to predict total energy, the differentiability provides forces that enable geometry optimization of large structures or simulate dynamics on longer timescales than traditionally feasible~\cite{Schtt2018}.

To train the models in this work, we use the first \num{10000} structures as the training dataset and the last \num{2500} structures to calculate the out-of-sample predictive performance. Finally, as negative gaps are unphysical, the models' predictions are clipped to zero, i.e. $\hat{y}=\max(0,f(\bm{x}))$.

\subsection{Kernel Ridge Regression with SOAP}
Kernel ridge regression (KRR) has been used for a variety of studies involving quantum-mechanical calculations~\cite{PhysRevLett.108.058301,Rupp2015,2015arXiv150307406F,De2016,1704.06439,Jger2018}.

KRR is based on linear ridge regression for the features $\bm{x}\in\mathcal{X}$ transformed to a higher-dimensional space $\mathcal{Z}$. It turns out that it is possible to perform ridge regression on the transformed datapoints $\bm{z}_i=\phi(\bm{x}_i)$ without explicitly converting the points by using the so-called kernel trick. This is achieved by defining a kernel that corresponds to an inner product in $\mathcal{Z}$, i.e. $k(\bm{x},\bm{x'})=\left<\phi(\bm{x}), \phi(\bm{x}')\right>$. For a derivation of KRR see~\cite{murphy2012machine}.

KRR introduces one constant parameter in the model that has to be chosen before training, i.e. a hyperparameter. This is the regularization coefficient $\lambda$. In this paper, it is optimized using a grid search within 10-fold cross validation (see Appendix for details). 

The goal of the SOAP kernel is to provide a measure of similarity between two structures, i.e. a structural similarity kernel~\cite{PhysRevB.87.184115,De2016}. In practice, this is a two-step process. First, a measure of similarity between two local environments $\mathcal{X}$ and $\mathcal{X'}$ is defined. Second, these local environment kernels are combined into a global similarity kernel. The local environment of an atom is constructed by placing a Gaussian at each neighboring atom position. The similarity (or overlap) between two environments $K(\mathcal{X},\mathcal{X}')$ is calculated by integrating the overlap under all SO(3) rotations. For more details on SOAP see~\cite{PhysRevB.87.184115,2018arXiv180700408W}.

The local environment kernels also depend on a number of hyperparameters. The most important is the cut-off distance $r_c$ that determines the radius of the environments considered. Additionally, the integration is in practice performed by expanding the environments in spherical and radial basis functions. This gives two additional hyperparameters, the number of radial basis functions $n$ and the maximum degree $l$ of the spherical harmonics. Figure~\ref{fig:soap_benchmark} and Table~\ref{tab:SOAP_performance} in the appendix show the performance for a number of different choices of $r_c$. The parameters $n=8$ and $l=6$ are chosen sufficiently high as was shown in \cite{PhysRevB.87.184115}.

There are several ways local environment kernels can be combined into a global similarity kernel. Two common choices are the fast average kernel and the ``regularized-entropy match'' (RE-Match) kernel \cite{De2016}. The fast average kernel corresponds to taking the average of all pair-wise environment kernels, whereas the RE-Match kernel is based on finding the best matches for each atomic environment \cite{De2016}. However, the RE-Match scheme is prohibitively slow and memory-consuming in its current implementations for the presented dataset. Hence, the fast average kernel is used, which implementation is available at \url{https://github.com/cosmo-epfl/glosim}.

\subsection{SchNet}
The SchNet model is a deep learning architecture based on artificial neural networks that is designed to work with both molecules and crystal structures~\cite{Schtt2018}. This deep learning model is ``end-to-end'' and does not require hand-crafted descriptors of the input data. Its architecture is inspired by the success of convolutional neural networks applied to image and video data. Other similar models include the HIP-NN~\cite{Lubbers2018} and the crystal graph convolutional networks \cite{PhysRevLett.120.145301}. 

In contrast to images, atomic positions cannot be efficiently represented using a simple uniform grid structure. Additionally, images usually have a fixed number of pixels whereas crystals can contain varying number of atoms. These issues are resolved by using the weight-sharing approach, which is common in both convolutional and recurrent neural networks. It assumes that the same network is used for each individual atom leading to an atom-wise contribution to the final prediction. The total number of contributions for a crystal structure is equal to the total number of atoms, and the final prediction is obtained by either averaging or summing all the contributions, referred to as ``pooling''. The pooling procedure maintains the atom index invariance because summing or averaging the atomic contributions is communicative. In practice, this means that SchNet decomposes a property into individual atomic contributions. For intensive properties, the mean of the atom contributions is taken, otherwise the sum is used. Besides introducing atom index invariance, it makes the model partially interpretable, since the final prediction can be explained by these individual contributions.

The SchNet architecture consists of four building blocks: atom embedding, atom-wise layers, interaction blocks and filter-generating networks. First, atoms are represented (embedded) by a vector $\bm{a}_{Z}\in\mathbb{R}^F$ depending on the proton number $Z$. $F$ is the dimensionality of the atom embedding. Interactions with neighbouring atoms (including periodic boundary conditions) are introduced by the interaction blocks. Here, a cut-off for the environments considered is used, which we take as \SI{5}{\angstrom} as in the original paper. The contributions by neighbouring environments in the interaction blocks are mediated by atom-wise layers and filter-generating networks. The computational details of these components are beyond the scope of this article but are available in \cite{Schtt2018,Schtt2018SchNetPack}.

The embeddings and atom-wise layers are optimized for the target property $y$ using a gradient descent method on the squared loss function $\min\limits_\theta L(\theta)=||y-\hat{y}||^2$, where $\theta$ are parameters of the network. Similar to \cite{Schtt2018SchNetPack}, we reduce the learning rate $\eta=0.001$ with a decay factor of $0.6$, and use the ADAM optimizer~\cite{DBLP:journals/corr/KingmaB14}. We train SchNet with a batch size of \num{32} and use a validation set of \num{1000} examples for early stopping or as long as $\eta'>\eta/10$. The latest implementation of SchNet in SchNetPack is used here, available at \url{https://github.com/atomistic-machine-learning/schnetpack}. We use two NVIDIA Tesla K80 for training.

Besides the standard training hyperparameters such as learning rate $\eta$ and batch size, there are a number of hyperparameters specific to SchNet. First, the cut-off radius $r_c$ is set to \SI{5}{\angstrom}, which was demonstrated to lead to accurate results on bulk crystal structures from the Materials Project~\cite{Schtt2018}. Second, the dimensionality of the embeddings $F$ is set to 64. Finally, the number of interaction blocks $T$ is varied between 3 and 6 (see Appendix~\ref{appendix:hp} and Figure~\ref{fig:schnet_benchmark} for more details).

\section{Hyperparameter optimization}\label{appendix:hp}

The machine learning models used for band gap prediction in this work, SOAP and SchNet, include parameters that have to be chosen before training begins, referred to as hyperparameters. The two models have different hyperparameters which are discussed separately in this appendix.

\subsection{SOAP}

\begin{figure}[h]
    \centering
    \includegraphics{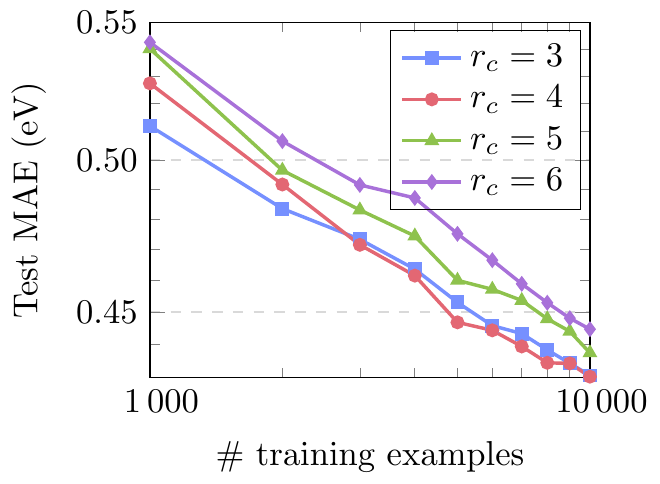}
    \caption{Performance of SOAP (average global kernel with $n=8$ and $l=6$) on the OMDB band gap dataset (\datasetname{}).}
    \label{fig:soap_benchmark}
\end{figure}

The \textit{Smooth Overlap of Atomic Positions} kernel relies on several hyperparameters. First, an environment $\rho$ (i.e. a set of neighboring atoms) is constructed by placing a single Gaussian on each neighboring atom position. The kernel $k(A,B)$ is calculated for two different environments A and B as the overlap of $\rho_A$ and $\rho_B$ while also integrating over all possible rotations~\cite{PhysRevB.87.184115}. The number of radial $n$ and the maximum degree of spherical harmonics $l$, determine the precision with which the overlap integral is performed. These parameters are set to $n=8,l=6$ as this was shown to lead to accurate results~\cite{PhysRevB.87.184115}. Finally, the most important hyperparameter is the cut-off radius $r_c$ that determines the cut-off radius for the local environments, i.e. which neighboring atom positions are included in the overlap integral. Figure~\ref{fig:soap_benchmark} shows the performance of SOAP for varying cut-off radii $r_c$. The lowest mean absolute error (MAE) is obtained for $r_c=\SI{4}{angstrom}$.
Table~\ref{tab:SOAP_performance} summarizes the performance of varying $r_c$ with all \num{10000} training examples.
\begin{table}[h]
    \centering
    \caption{Summary of SOAP (with average kernel $n=8$ and $l=6$) performance with \num{10000} training examples.}
    \begin{tabular}{ccc}
        $r_c$ (\AA) & MAE (eV) & RMSE (eV)\\
        \hline
        3 & 0.431 & 0.582\\
        \textbf{4} & \textbf{0.430} & \textbf{0.576}\\
        5 & 0.437 & 0.594\\
        6 & 0.445 & 0.601\\
    \end{tabular}
    \label{tab:SOAP_performance}
\end{table}

\subsection{SchNet}

\begin{figure}[h]
    \centering
    \includegraphics[]{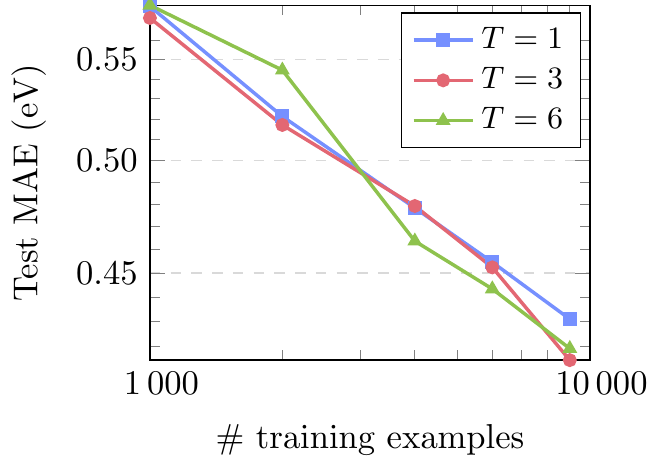}
    \caption{Benchmark of SchNet on the OMDB band gap dataset (\datasetname{}). $F=64$,$r_c=\SI{5}{\angstrom}$}
    \label{fig:schnet_benchmark}
\end{figure}

The SchNet model is an artificial neural network that consists of four different components: atom embeddings, atom-wise layers, interaction blocks and filter-generating networks. See \cite{Schtt2018} for a schematic drawing of the model and its components. The interactions with neighboring atoms (including periodic boundary conditions) are included through interaction blocks that rely on filter generators. Increasing the number interaction blocks $T$ in the model increases the depth of the model and the number of parameters $\theta$. For example, a model with $F=64$ and $T=3$ has \num{63333} parameters, whereas a model with $T=6$ has \num{117953} parameters.  $T$ is a hyperparameter to be set before training. SchNet is evaluated on each site in the crystal structure and the final band gap prediction is the mean of the atom-wise contributions. Figure~\ref{fig:schnet_benchmark} show the performance of SchNet for a varying number of interactions blocks $T$ and training examples. Table~\ref{tab:SchNet_performance} summerizes the results with the maximum number of 9000 training examples (the remaining 1000 validation examples are used for model selection).
\begin{table}[h]
    \centering
    \caption{Summary of SchNet performance with \num{9000} training examples with \num{1000} validation examples for model selection (early stopping), $F=64$, $r_c=\SI{5}{\angstrom}$.}
    \begin{tabular}{ccc}
        $T$ & MAE (eV) & RMSE (eV)\\
        \hline
        1 & 0.431 & 0.584\\
        \textbf{3} & \textbf{0.415} & \textbf{0.554}\\
        6 & 0.419 & 0.565\\
    \end{tabular}
    \label{tab:SchNet_performance}
\end{table}

\begin{figure}[h!]
    \centering
    \includegraphics[]{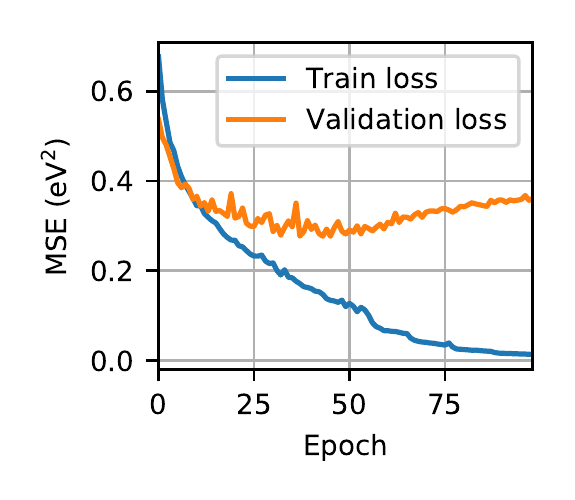}
    \caption{The training and validation loss while training SchNet with 9000 training examples and $T=3$.}
    \label{fig:schnet_curves}
\end{figure}

Figure~\ref{fig:schnet_curves} shows the training and validation loss during the training of SchNet (with $T=3$). After many epochs, SchNet is able to overfit to the training data completely due to its many free parameters. The validation loss is used to select the best model, which in this case occurs at epoch 44.

\begin{figure}[h!]
	\centering
	\includegraphics[width=\linewidth]{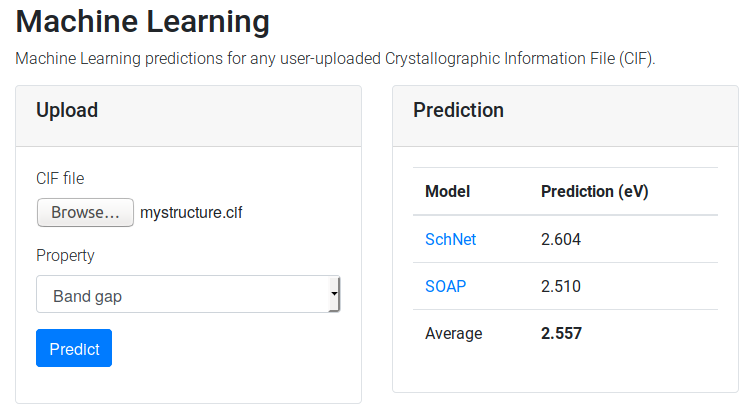}
	\caption{The trained models are available through a web interface at \url{https://omdb.mathub.io/ml}.}
	\label{fig:OMDB_ML_service}
\end{figure}

\section{Web interface}
\label{appendix:web_interface}
We implemented a web interface that can predict band gap properties for a user-uploaded Crystallographic Information File (CIF) containing the crystal structure (Figure~\ref{fig:OMDB_ML_service}). The trained models scale differently when presented a new crystal structure. SchNet is a parametric model with a fixed number of parameters per atom that is evaluated on each site. Therefore, it scales linearly with the the number of atoms in the unit cell \cite{Schtt2018}. Kernel ridge regression, on the other hand, scales with the number of training samples $N$ because it requires calculating the kernel-based similarity between the new structure $A$ and all the reference examples $B_i$ in the training set to make a prediction of the target property $g$, i.e. $g=\sum_i^N \alpha_i K(A,B_i)$, where $\alpha_i$ are the regression coefficients. It takes $\approx$ 10 seconds for SchNet and $\approx$ 1 minute for SOAP to calculate the prediction using the online interface.

\bibliography{references}

\end{document}